\documentclass[10pt,twocolumn,letterpaper]{article}

\usepackage{wacv}
\usepackage{times}
\usepackage{epsfig}
\usepackage{graphicx}
\usepackage{amsmath}
\usepackage{amssymb}
\usepackage{algorithm}
\usepackage{algorithmic}
\usepackage{verbatim}
\usepackage{subfigure}
\usepackage{float}
\usepackage{authblk}

\def\wacvPaperID{1058} 

\wacvfinalcopy 

\ifwacvfinal
\def\assignedStartPage{1} 
\fi

\ifwacvfinal
\usepackage[breaklinks=true,bookmarks=false]{hyperref}
\else
\usepackage[pagebackref=true,breaklinks=true,colorlinks,bookmarks=false]{hyperref}
\fi

\ifwacvfinal
\else
\fi

\begin{document}

\title{``Sliced" Subwindow Search: a Sublinear-complexity\\ Solution to the Maximum Rectangle Problem}

\author[1,2]{Max Reuter\thanks{reuterm1@msu.edu}}
\author[1]{Gheorghe-Teodor Bercea\thanks{gheorghe-teod.bercea@ibm.com}}
\author[1]{Liana Fong}
\affil[1]{IBM T.J. Watson Research Center, Yorktown Heights, NY}
\affil[2]{Department of Computer Science and Engineering, Michigan State University}

\renewcommand\Authands{, }

\maketitle
\thispagestyle{empty}

\begin{abstract}
Considering a 2D matrix of positive and negative numbers, how might one draw a rectangle within it whose contents sum higher than all other rectangles'? This fundamental problem, commonly known the maximum rectangle problem or subwindow search, spans many computational domains. Yet, the problem has not been solved without demanding computational resources at least linearly proportional to the size of the matrix. In this work, we present a new approach to the problem which achieves sublinear time and memory complexities by interpolating between a small amount of equidistant sections of the matrix. Applied to natural images, our solution outperforms the state-of-the-art by achieving an 11x increase in speed and memory efficiency at 99\% comparative accuracy. In general, our solution outperforms existing solutions when matrices are sufficiently large and a marginal decrease in accuracy is acceptable, such as in many problems involving natural images. As such, it is well-suited for real-time application and in a variety of computationally hard instances of the maximum rectangle problem.
\end{abstract}

\section{Introduction}

The problem of locating the maximal-sum rectangle in a 2D matrix spans a variety of disciplines, including computer vision~\cite{an2009efficient, lampert2008beyond}, data mining~\cite{fukuda1996data}, data analysis~\cite{eckstein2002maximum}, pattern matching~\cite{fischer1993approximations}, computational theory~\cite{backurs2016tight}, and more~\cite{agarwal2006hunting, backer2010mono, liu2003planar}. Known as the max-weight rectangle problem in its purest form, the problem's definition considers $n \times m$ matrices. As in many existing works, we will instead consider inputs of $n \times n$ matrices without much loss of generality in practice, with \mbox{$O(N) = O(n^2)$} as the definition of linear complexity.

\begin{figure}
\centering
\includegraphics[scale=0.6]{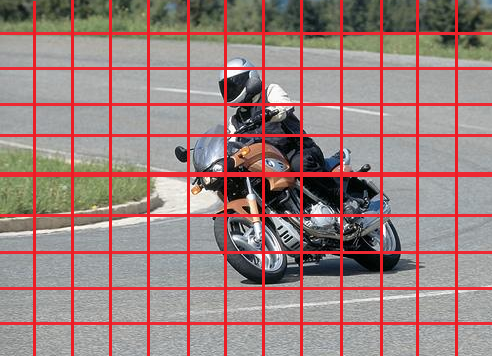}
\caption{Matrix slices used to guide \textbf{Sliced Subwindow Search}, the number of which are determined by a user-chosen function of the matrix size. Only this subset of the matrix is read during search, drastically reducing running time and memory footprint.}
\label{fig:Sampling}
\end{figure}

Despite decades of its study, even approximate solutions to the max-weight rectangle problem require substantial computational resources~\cite{an2009efficient, bentley1984programming, lampert2008beyond, backurs2016tight}. The problem's leading and long-standing solutions are comprised of an exact algorithm which runs in \mbox{$O(n^3)$} time~\cite{bentley1984programming}, an exact algorithm which runs in \mbox{$O(n^2)$} expected time to \mbox{$O(n^4)$} worst-case time~\cite{lampert2008beyond}, and an approximate algorithm which runs in \mbox{$O(n^2)$} expected time to \mbox{$O(n^3)$} worst-case time~\cite{an2009efficient}. In this work, we will present an algorithm which achieves sublinear time: \mbox{$O(\frac{n^2}{f(n)})$}, with $f(n)$ a user-chosen function of the matrix's size (see Figure~\ref{fig:Sampling}).

\subsection{Subwindow Search for Object Localization} \label{Object localization}
Object localization invokes the maximum rectangle problem by extracting image features into a 2D matrix of the same dimensions as the image, corresponding to ``pixels of interest". Then, subwindow search is used to locate a highly positive section of this feature matrix, yielding a bounding box around the ``area of most interest". Critically, in all current solutions to the maximum rectangle problem, any similarity among nearby elements of the input matrix is not exploited. In this work, we focus on the problem of object localization, which allows for a clear and highly practical demonstration of the advantage of doing so.

\subsection{Contributions}
We present \mbox{\textbf{Sliced Subwindow Search}} (\textbf{SSS}), an approximation algorithm for the max-weight rectangle problem which scales considerably more gracefully than existing solutions (see Table~\ref{tab:1}). Our algorithm leverages similarity among nearby matrix values via equidistant sampling, allowing for inferences to be made about the unsampled data without dedicating additional computational resources to it. We also show advantageous performance of our algorithm on sufficiently large matrices which exhibit no nearby-element similarity. While sampling is a common technique used to trade accuracy for speed across many problems, our novelty contribution lies in the non-trivial embedding of sampling as a concept within the max-weight rectangle problem's optimal subroutines.

In summary, we make the following contributions:
\begin{itemize}
\itemsep0em 
\item We present an approximate solution to the maximum rectangle problem which runs in sublinear time and memory, significantly reducing running time and memory usage compared to all existing solutions.
\item We evaluate our approach on the PASCAL VOC 2006 dataset and obtain an 11x increase in speed and memory efficiency over the state-of-the-art solution with 99\% comparative accuracy.
\item We demonstrate that an increase in either the size or the nearby-element similarity of input matrices rapidly increases the relative performance of our solution.
\end{itemize}

\begin{table}
\begin{center}
\begin{tabular}{|c|c|c|c|}
\hline
Algorithm & Accuracy & Time & Memory\\
\hline\hline
Bentley & Exact & $O(n^3)$ & $O(n^2)$ \\
ESS & Exact & $O(n^2)$ & $O(n^2)$ \\
\mbox{A-ESS} & Approximate & $O(n^2)$ & $O(n^2)$ \\
\mbox{\textbf{SSS}} & Approximate & $O(\frac{n^2}{f(n)})$ & $O(\frac{n^2}{f(n)})$ \\
\hline
\end{tabular}
\end{center}
\caption{Expected complexities for the max-weight rectangle problem on \mbox{$n \times n$} matrices. Note that \mbox{A-ESS} runs 900x faster than ESS on average in practice~\cite{an2009efficient}.}
\label{tab:1}
\end{table}

\section{Related Works}\label{Problem overview}
Formally, for an \mbox{$n \times n$} matrix of positive and negative numbers, its maximum submatrix is that whose sum is largest among that of all possible submatrices. That is, for some \mbox{$n \times n$} matrix $M$ with x-axis indices \mbox{$0 \leq x < n$} and y-axis indices \mbox{$0 \leq y < n$}, the maximum submatrix with maximum sum \mbox{$S_{max}$} and optimal bounds \mbox{$[(x_1,y_1),(x_2,y_2)]$} with \mbox{$x_1 \leq x_2$ and $y_1 \leq y_2$} is defined as:

\begin{equation} \label{eq:1}
S_{max} \triangleq \sum\limits_{x=x_1}^{x_2}\sum\limits_{y=y_1}^{y_2} M[x, y]
\end{equation}

The max-weight rectangle problem can be trivially brute-forced by examining all possible \mbox{$n^4$} submatrices and maximizing across them in \mbox{$O(n^6)$} time ($O(n^2)$ spent on each to sum their contents). This solution uses \mbox{$O(1)$} space and yields the globally optimal submatrix. In contrast to this solution, an approximation algorithm simply \emph{attempts} to yield a globally optimal submatrix. In this work, we measure the quality of a given solution's approximated submatrix by calculating its area of overlap (more precisely, the intersection over union or IoU) with the global optimum.

\subsection{Kadane's Algorithm: a Key Subroutine}
At the core of all prominent solutions to the max-weight rectangle problem is Kadane's algorithm (1977)~\cite{bentley1984programming}, a solution to the 1D maximum-subarray problem: a search for the subarray within an array whose sum is largest among that of all subarrays. Formally, for some array $A$ of size $n$, its maximum subarray with maximum sum $S_{max}$ and optimal bounds \mbox{$[x_1, x_2]$} for \mbox{$0 \leq x < n$ with $x_1 \leq x_2$} is defined as:

\begin{equation} \label{eq:2}
S_{max} \triangleq \sum\limits_{x=x_1}^{x_2}A[x]
\end{equation}

Kadane's algorithm is implemented as follows. First, it traverses the input array, collecting array elements until the running sum of the collection is negative, at which point it discards the collection and begins a new one. While traversing, it records intervals of the array corresponding to a collection whose sum is maximal among those it has seen thus far. Ultimately, when it reaches the end of the array, its record at that time describes the collection which had the globally maximum sum. It then returns the bounds and sum of that collection.

As Kadane's algorithm performs a constant number of calculations for each array element, its running time corresponds to the amount of time taken to traverse the array: \mbox{$O(n)$}. It uses space to store the running maximum sum and its boundaries, both of which are independent of the input size. Thus, its space complexity is \mbox{$O(1)$}.

\subsection{Bentley's Algorithm: an Exact Solution}
Bentley's algorithm (1984)~\cite{bentley1984programming} provides a \mbox{$O(n^3)$} time and \mbox{$O(n^2)$} space exact algorithm for the max-weight rectangle problem. It does so by examining all possible column-to-column pairs and aggregating the values between them into an array, which is then sent to Kadane's algorithm. The returned column combination \mbox{$[x'_1,x'_2]$} with the maximum sum then describes the horizontal bounds of the globally optimal submatrix. Then, the optimal vertical \mbox{$[y'_1,y'_2]$} bounds are determined by a similar aggregation and maximization, \emph{but within these horizontal bounds}.

As there are \mbox{$O(n^2)$ $[x_1,x_2]$} column-to-column combinations in a given matrix, and applying Kadane's algorithm to each corresponding 1D aggregation costs $O(n)$ time, the running time of Bentley's algorithm is \mbox{$O(n^2) \times O(n) = O(n^3)$}. Crucially, the aggregation of a given column-to-column combination into a 1D array must be constant time to achieve this complexity. This can be accomplished with preprocessing: first computing the prefix sums of the matrix, allowing the sum between two points to be computed in \mbox{$O(1)$} time. This preprocessing takes \mbox{$O(n^2)$} time and occupies \mbox{$O(n^2)$} space, determining the algorithm's \mbox{$O(n^2)$} space complexity.

\subsection{Efficient Subwindow Search (ESS)}
Efficient Subwindow Search (ESS) (2008)~\cite{lampert2008beyond} is another exact solution which has utility in object localization by maximizing across a large class of classifier functions over all subimages. A branch-and-bound method, ESS guarantees a globally optimal solution in \mbox{$O(n^2)$} expected time to \mbox{$O(n^4)$} worst-case time using \mbox{$O(n^2)$} expected space to \mbox{$O(n^4)$} worst-case space. Its worst-case time on the PASCAL VOC 2006 dataset is measured to take 52x more time to compute than the average case of Bentley's \mbox{$O(n^3)$} algorithm, and on average its running time on the dataset is 1.1x that of Bentley's algorithm~\cite{an2009efficient}, though it is more practical than Bentley's algorithm in different object detection and retrieval scenarios.

ESS examines disjoint subsets of the set of all possible submatrices in a best-first manner, noting the maximal quality in each subset. This allows for the discarding of many subset search spaces due to their maximal qualities being inadequate with respect to other previous states. The algorithm calculates integral images in preprocessing, costing \mbox{$O(n^2)$} time and \mbox{$O(n^2)$} space. In the worst case of ESS, all submatrices must be examined, of which there are \mbox{$O(n^4)$}, resulting in the algorithm's worst-case time and space complexities.

\subsection{Alternating Efficient Subwindow Search \mbox{(A-ESS)}}
\begin{figure}
\centering
\includegraphics[scale=0.375]{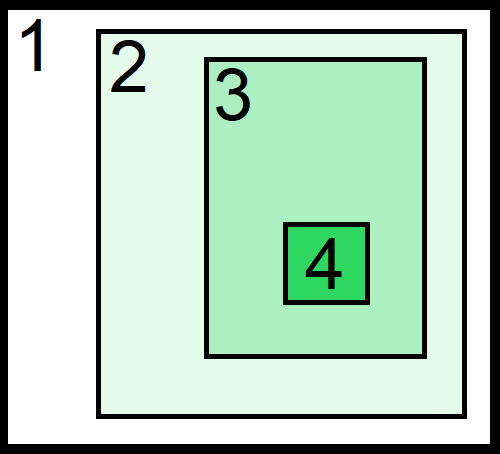}
\caption{Iterative refinement of bounds conducted by \mbox{A-ESS} and \mbox{SSS}. Note that not each iteration's bounding box need be strictly within bounds of the that of the previous iteration.}
\label{fig:Iterative Search}
\end{figure}

Alternating Efficient Subwindow Search \mbox{(A-ESS)} (2009)~\cite{an2009efficient} is an approximation algorithm that runs in \mbox{$O(n^2)$} expected time using \mbox{$O(n^2)$} space, though its worst-case time is \mbox{$O(n^3)$}. Like Bentley's algorithm, \mbox{A-ESS} begins by computing the horizontal and vertical prefix sums of the input matrix, requiring \mbox{$O(n^2)$} time and space. It then initializes a candidate submatrix the size of the matrix itself. At each iteration, values \emph{along} the horizontal bounds (not exclusively those within both the horizontal and vertical bounds) are aggregated into a 1D array. Similarly, values along the vertical bounds are aggregated. Then, both of these 1D arrays are evaluated by Kadane's algorithm, using the returned bounds for the candidate submatrix in the next iteration. This iterative refinement process continues until a further iteration would cause a decrease in the sum of the values within the candidate bounding box (see Figure~\ref{fig:Iterative Search}).

The score function used to determine the quality of a candidate bounding box and whether another iteration should take place is the difference between the maximum sums of the current iteration's 1D horizontal and vertical subarrays. This function is non-decreasing, and the optimal score is bounded by the sum of all the matrix's positive elements. Thus, the algorithm is convergent.

On the PASCAL VOC 2006 natural image dataset, it has been shown that the number of iterations in A-ESS' search loop is typically 3-6~\cite{an2009efficient}, thus the expected computational complexity is determined by the upper bound of a constant number of loop iterations and any preprocessing. A loop iteration costs only \mbox{$O(n)$} time and \mbox{$O(n)$} space to aggregate values into 1D arrays, leveraging the prefix sum matrices. However, calculating the prefix sum matrices costs \mbox{$O(n^2)$} time and \mbox{$O(n^2)$} space, thus the algorithm's time complexity is typically dominated by this preprocessing. Finally, because there are \mbox{$O(n^2)$} possible row and column intervals to search and an iteration costs \mbox{$O(n)$} time, the algorithm's worst-case time complexity is \mbox{$O(n^2) \times O(n) = O(n^3)$}. As \mbox{A-ESS} is exceedingly effective in practice, we consider this the current state-of-the-art algorithm.

\subsection{Other Solutions}
An exact algorithm exists which is most useful when applied to large matrices with time complexity \mbox{$O(n^3(\frac{\log \log n} {\log n})^{1/2}) < O(n^3)$}~\cite{tamaki1998algorithms}. However, matrices with \mbox{$n < 1,000,000$} cause the algorithm to perform less efficiently than Bentley's algorithm due to preprocessing overheads~\cite{an2009efficient}, and so it is commonly impractical.

A distributed exact solution of the problem's workload exists that allows it to be solved in \mbox{$O(\log n)$} time across \mbox{$O(\frac{n^3}{\log n})$} workers~\cite{wen1995fast}. However, on the PASCAL VOC 2006 dataset, achieving this \mbox{$O(\log n)$} running time requires millions of workers, and achieving a running time competitive with ESS or \mbox{A-ESS} (on the order of \mbox{$O(n^2)$}) would still require dozens of workers.

\section{Sliced Subwindow Search \mbox{(SSS)}}\label{Sliced Subwindow Search (SSS)}
Our approach, which we call Sliced Subwindow Search or \mbox{SSS}, is an approximation algorithm that partially samples a given matrix rather than examining all of its elements (see Figure~\ref{fig:Sampling}). The algorithm is similar to \mbox{A-ESS} in that it initializes a candidate bounding box that is the size of the matrix itself, then iteratively refines its bounds towards more desirable submatrices. However, it computes only parts of the horizontal and vertical prefix sums, computing only prefix sums of equidistant rows and columns at a stride length of some user-chosen \mbox{$f(n)$}. For example, given a matrix with 100 rows, a chosen \mbox{$f(n) = \sqrt{n} = 10$}, and an offset of \mbox{$\frac{f(n)}{2} = 5$}, rows \mbox{$5, 15, 25, \dots 85, 95$} will have their horizontal prefix sums computed. Similarly, equidistant columns will have their prefix sums computed. The other rows and columns can be omitted as they will not be referenced during the search loop. Thus, the prefix sums occupy \mbox{$O(\frac{n^2}{f(n)})$} space rather than \mbox{A-ESS'} \mbox{$O(n^2)$}, determining the two algorithms' difference in memory complexity.

After the prefix sums have been computed, SSS initializes the candidate bounding box that is the size of the matrix itself, then begins a search loop similar to that of \mbox{A-ESS}'. First, it aggregates the columns within range \mbox{$[x_1,x_2]$} into a 1D vertically-oriented array via summation. However, in contrast to \mbox{A-ESS}, not all columns are selected for aggregation. Only equidistant sections of the columns \mbox{$f(n)$} rows apart are written to the 1D array, referencing the previously computed partial horizontal prefix sum matrix, and the rest of the 1D array is populated with zeros. For example, given a matrix with 100 rows and \mbox{$f(n) = \sqrt{n} = 10$} and an offset of \mbox{$\frac{f(n)}{2} = 5$}, indices \mbox{$5, 15, 25, \dots 85, 95$} of columns within range \mbox{$[x_1,x_2]$} are sampled and have their sums written to the 1D array at indices \mbox{$5, 15, 25, \dots 85, 95$}, respectively. Then, the 1D array with zeros at the indices where sampling was not done is sent to Kadane's algorithm which returns optimal bounds that serve as top and bottom boundaries \mbox{$[y_1',y_2']$} for the next iteration of the candidate submatrix. A similar process is repeated for the rows, obtaining new left and right \mbox{$[x_1',x_2']$} bounds using the partial vertical prefix sum matrix. \mbox{$[x_1',x_2']$} and \mbox{$[y_1',y_2']$} are then used to define the new candidate bounding box's boundaries, and the iterative refinement process continues. The previous iteration's maximum sum of the partially populated 1D array fed to Kadane's algorithm is compared to the current iteration's, and the program returns when the current iteration's sum is the smaller of the two values, indicating a likely decrease in the candidate bounding box's quality. See Algorithm~\ref{alg:Code} for the algorithm's pseudocode.

The search space contains \mbox{$O(n^2)$} points, so if it is sampled every \mbox{$f(n)$} rows or columns via striding, \mbox{$O(\frac{n^2}{f(n)})$} points will be examined. Given this and the employment of an iteration cap $I_{cap}$ (see Section~\ref{sec:accuracy-determinants}), the algorithm runs in \mbox{$O(\frac{n^2}{f(n)})$} time. Thus, if the size of the input $O(N)=O(n^2)$ is considered as the reference point for linearity, \mbox{SSS} is sublinear in its time and memory usage.

\begin{algorithm}[t!]
\caption{Sliced Subwindow Search (SSS)}
\label{alg:Code}
\hspace*{\algorithmicindent}Require: $n \times n$ matrix $M$\\
\hspace*{\algorithmicindent}Preprocessing: compute partial horizontal and prefix \hspace*{3mm} sums $P_h$ and $P_v$;\\
\hspace*{\algorithmicindent}Initialization: Set $(x_1,x_2)=(y_1,y_2)=(1,n)$, $g = 1,$
\hspace*{\algorithmicindent}$I = 1;$

\begin{algorithmic}
\WHILE{$g > 0 \text{ }\AND\text{ } I \leq I_{cap}$}
\STATE 1. Populate arrays $a$ and $b$ with zeros;
\STATE 2. Compute $a[j] = \sum\limits_{i=y_1}^{y_2} M[i,j]$ with $j$ incrementing \hspace*{3mm} by $f(n)$ from $1$ to $n$ using $P_v$;
\STATE 3. Apply Kadane's algorithm to $a$ and find the optimal \hspace*{3mm} column interval $[x_1',x_2']$ and the maximum $s_1$;
\STATE 4. Compute $b[i] = \sum\limits_{j=x_1'}^{x_2'} M[i,j]$ with $i$ incrementing \hspace*{3mm} by $f(n)$ from $1$ to $n$ using $P_h$;
\STATE 5. Apply Kadane's algorithm to $b$ and find the \hspace*{3mm} \mbox{optimal} row interval $[y_1',y_2']$ and the maximum $s$;
\STATE 6. Set the gain $g = s - s_1$;
\STATE 7. Increment the iteration count $I = I + 1$;
\ENDWHILE
\end{algorithmic}
\end{algorithm}

\subsection{Convergence}
\label{sec:convergence}
Unlike \mbox{A-ESS}, \mbox{SSS} is not guaranteed to converge without the usage of an iteration cap due to its use of sampling. However, similar to \mbox{A-ESS}, the number of iterations needed to reach a high-quality bounding box is generally low. The looping case can be triggered by two edge cases: (1) the search is narrowed down to a 1-dimensional array that is not represented in the sampling, or (2) missing data due to sampling causes the horizontal and vertical searches to suggest to each other bounds whose contents the other cannot completely evaluate in a particularly problematic way. This can cause the objective function to falsely report that a search iteration was profitable and that the search should continue. We thus use $I_{cap}$ to limit the number of iterations performed during the search phase. Cases where this cap is reached result in very low accuracy, but are most common among small and random matrices, and their occurrence drops precipitously as matrices become larger and/or more spatially coherent.

\subsection{Accuracy}
\label{sec:accuracy-determinants}
Theoretical bounds and guarantees of \mbox{SSS} have limited practical use due to its data-dependent nature (partially inherited from \mbox{A-ESS}). In the best case, the sampled data is sufficient to inform a perfect, globally optimal bounding box (100\% accuracy). In the worst case, the sampled data is insufficient (such as no meaningful data being sampled), resulting in a low-quality bounding box proposal (0\% accuracy). Therefore, in general, no guarantees can be supplied, and certainly not without knowledge of the nature of the input data. In light of this, we provide a spatial coherence score (see Section~\ref{subsec:coherence}) measuring the patterned nature of the input that, in our experiments across many types of data, is a strong predictor of \mbox{SSS'} accuracy.

\section{Performance Evaluation}
\label{Performance Evaluation}

\subsection{Experimental Setup}
We first compare the runtime and accuracy of \mbox{SSS} to that of \mbox{A-ESS} on the localization of 10 object classes in the PASCAL VOC 2006 dataset on a Power 9 CPU, compiling our code with Clang version 8.0. The experiments were executed with exclusive access to the compute node. To ensure a fair comparison, we use the exact same \mbox{A-ESS} implementation as was used in its publication~\cite{an2009efficient}, which we obtained directly from the authors. We also compare \mbox{SSS} to \mbox{A-ESS} on the RGB color channels of (1) digital art, \mbox{(2) a diverse} set of high-resolution images, and (3) randomized matrices of various sizes.

\subsection{Input Details}
\subsubsection{PASCAL VOC 2006}
The images in the PASCAL dataset contain around 200,000 pixels. Object features of the images and are represented as sparse matrices which contain around \mbox{$n^2 = 16,000$} entries. The dataset contains 5,304 images with 10 object classes each, resulting in a sample size of 53,040 bounding box searches.

\begin{figure}
\centering
\includegraphics[scale=0.38]{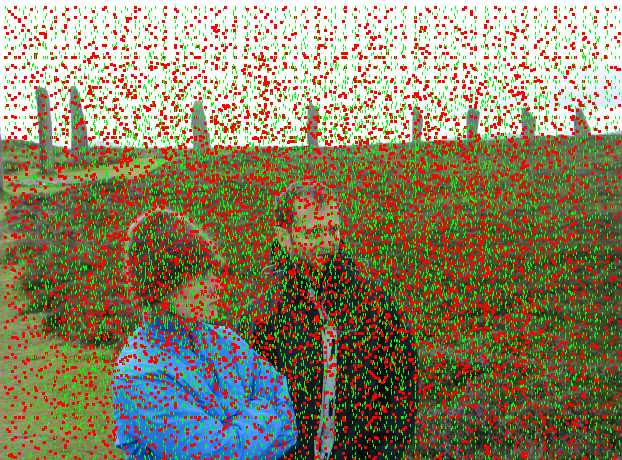}
\caption{A sparse object feature matrix with points sampled by \mbox{\textbf{SSS}} in green, and all non-sampled points in red.}
\label{fig:feature_strides}
\end{figure}

\begin{figure}
\centering
\includegraphics[scale=0.1865]{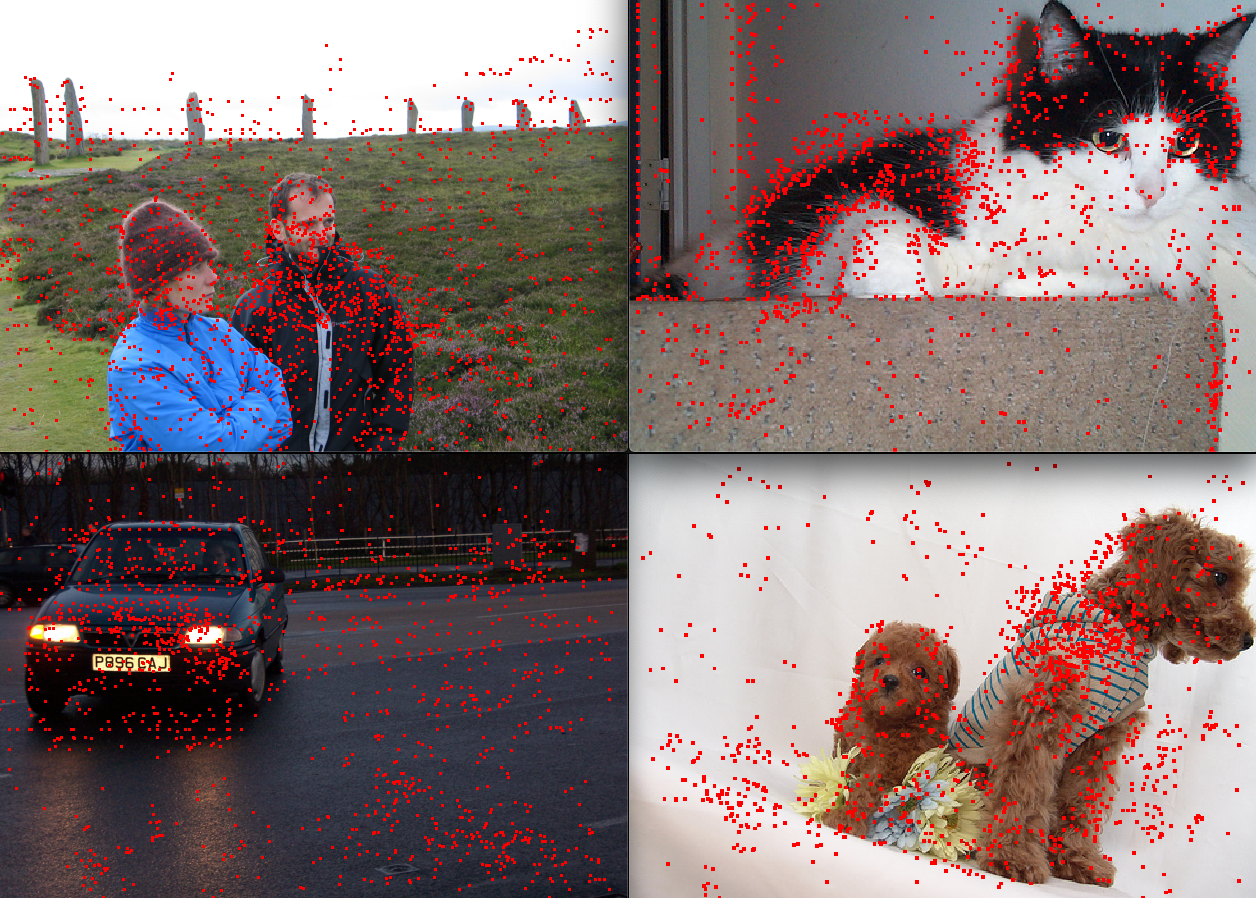}
\caption{Contents of sparse object feature matrices with only highly-positive points drawn.}
\label{fig:features}
\end{figure}

Entries in the object feature matrices are referred to as Sped Up Robust Feature (SURF) descriptors extracted from the original image using the same Support Vector Machine (SVM) formulation as in~\cite{blaschko2008learning}. Figure~\ref{fig:feature_strides} illustrates how a sparse matrix encodes an image and how \mbox{SSS} samples said sparse matrix. Entries in the sparse object feature matrix are attributed different positive or negative weights. Figure~\ref{fig:features} illustrates the highest weighted points in different images. These are points where the detector indicated increased likelihood of the object's presence.

To explore larger inputs, we increase the size of input sparse matrices using entry duplication. Each entry of the original matrix is transformed into a $K \times K$ submatrix with all entries equal to the original entry value. In this way, we arrive at a matrix that is $K^2$ times larger than the original matrix. In our experiments, we use $K \in \{1,2,3\}$.

\subsubsection{RGB Color Channels}
The RGB color channel data contains three types of data: 512x512 digital art, 1024x1024 digital art, and 2048x2048 high-resolution images, all obtained from the top 200 results of querying Google Images for ``digital art" and ``image", respectively, with the appropriate size constraints applied. Performance on randomized data of these sizes is also presented. For all of these datasets, the matrix entries are normalized such that the average entry is zero.

\subsection{Metrics}

We define the quality of an output submatrix, hereafter referred to as its accuracy, as a binary correct/incorrect evaluation on whether the output has an intersection over union (IoU) with the global optimum given some threshold, such as 50\%. On the PASCAL dataset, \mbox{A-ESS'} accuracy with respect to the global optimum is 96.3\% on all scaled versions of the data, meaning, for example, a 95\% accuracy of \mbox{SSS} with respect to \mbox{A-ESS} corresponds to 91.5\% global accuracy. On full (non-sparse) matrices, \mbox{A-ESS'} accuracy is between 98\% and 100\% and is thus a generally reliable reference point. We mostly use this reference point instead of computing the global optimum of all matrices in the given datasets, which would be very computationally expensive. We will henceforth use the term ``accuracy" to mean IoU with respect to \mbox{A-ESS}'s bounding boxes unless otherwise stated.

\subsubsection{Spatial Coherence Score}
\label{subsec:coherence}
We define a spatial coherence score -- how similar neighboring entries of a given matrix are -- to gain insight into \mbox{SSS'} accuracy across a variety of data. This score is obtained by calculating the average differences between matrix entries and their spatially neighboring entries. This results in a low score when a given matrix is noisy or random and a high score when a given matrix has many entries whose nearby entries are similar, such as one that contains a large blue sky. This type of spatial coherence also manifests in other patterns, such as objects (many ``dog-like" features are surrounded by similarly dog-like features).

Formally, we define the spatial coherence $C$ of a matrix $M$ as:
\begin{equation}
C = 1 - D
\end{equation}
where $D$ is the spatial dissimilarity in $M$, defined as
\begin{equation}
D = 
\frac{\sum\limits_{x=0}^{n}
\sum\limits_{y=0}^{n}
\sum\limits_{d_x=-r}^{r}
\sum\limits_{d_y=-r}^{r}
|M[x, y] - M[x + d_x, y + d_y]|}{c \times |M_{max} - M_{min}|}
\end{equation}
where $r$ is the square radius wherein a given entry of $M$ is compared to its neighbors, $M_{max}$ and $M_{min}$ are the maximum and minimum value entries in $M$, and $c$ is the total number of comparisons made. The denominator regularizes the entry differences aggregated in the numerator against the number of comparisons made, yielding a value \mbox{$D\in[0,1]\subset\mathbb{R}$}. $C = 1 - D$ then describes the inverse of the amount of spatial dissimilarity: the spatial coherence of the matrix. $C$ is thus a standardized metric for measuring and comparing the spatial coherences of matrices independent of their size, sparsity, and the magnitude of their entries.

\subsection{Independent Variables}
\label{subsec:variables}
In our experiments, we examine the effects on \mbox{SSS'} performance of five independent variables: (1) size of input $O(n^2)$, (2) data source, (3) IoU threshold, (4) randomization (on/off), and (5) \mbox{SSS'} stride lengths as functions $f(n)$ of the input size $O(n^2)$ in the following set:

\begin{equation} \label{eq:3}
\{\log \log n, \log n, \sqrt{n}, \log^2 n\}
\end{equation}

We use a constant iteration cap $I_{cap}$ of 20 for \mbox{SSS} when handling full matrices (no \mbox{PASCAL} trials triggered loop cases). We calculate the spatial coherence $C$ of a matrix using a square radius $r=5$ which we find to be sufficiently informative for the purposes of our analysis.

\subsection{Experiments}

\begin{figure*}
\centering
\includegraphics[scale=1]{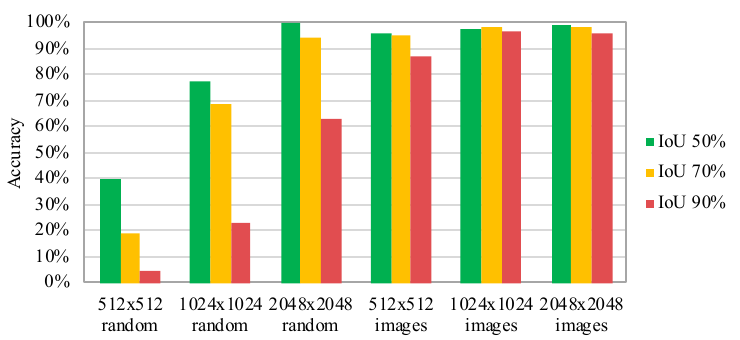}
\caption{\mbox{\textbf{SSS'}} accuracy on various data for various IoU percentages with \mbox{$f(n)=\log{n}$}.}
\label{fig:IoUs}
\end{figure*}

\begin{table}
\begin{center}
\begin{tabular}{|r|r|r|r|}
\hline
Scaling $K \times K$ & Bentley & A-ESS & \textbf{SSS}\\
 & [ms] & [ms] & [ms]\\
\hline\hline
(no scaling) 1 & 309 & 0.498 & 0.264\\
4 & 3,533 & 1.319 & 0.306\\
9 & 14,381 & 2.896 & \textbf{0.500}\\
\hline
\end{tabular}
\end{center}
\caption{Average running times of Bentley's algorithm, A-ESS, and \textbf{SSS} on PASCAL data. Times listed for \mbox{\textbf{SSS}} are those where the algorithm obtains at least 95\% accuracy with respect to \mbox{A-ESS}.}
\label{tab:Running Times}
\end{table}

\begin{figure}
\centering
\includegraphics[scale=1]{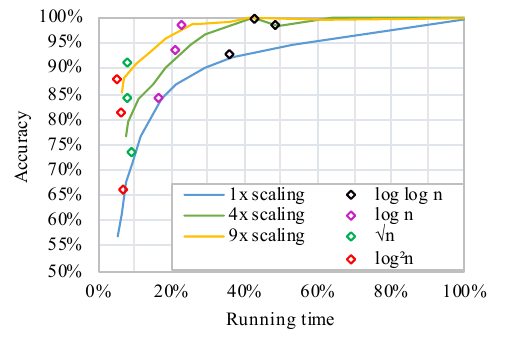}
\caption{Performance of \textbf{SSS} with respect to that of \mbox{A-ESS} on the \mbox{PASCAL} VOC 2006 dataset. Curve data between diamond points is obtained using constant values, such as $f(n) = 5$. Some diamond points deviate from the curves due to slight variance in image dimensions, an artifact of the generalization to $O(n^2)$ matrices as opposed to $O(n \times m)$.}
\label{fig:Experiment Results Functions}
\end{figure}

\begin{figure}
\centering
\includegraphics[scale=1]{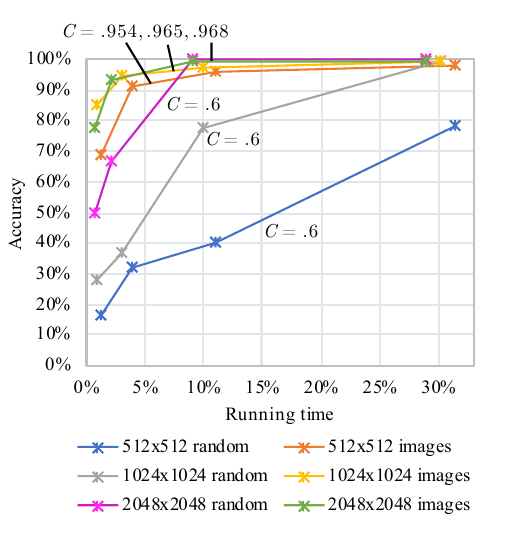}
\caption{\textbf{SSS'} performance with respect to that of A-ESS on 512x512 and 1024x1024 digital art, 2048x2048 high-resolution images, and correspondingly-sized randomized matrices for \mbox{$f(n) = \{\log \log n, \log n, \sqrt{n}, \log^2 n\}$} (from left to right). $C$ denotes a dataset's coherence score. \mbox{A-ESS} achieves \mbox{98\%-100\%} accuracy on all datasets.}
\label{fig:non-PASCAL}
\end{figure}

\mbox{Table~\ref{tab:Running Times}} presents the real-time running times of Bentley's algorithm, \mbox{A-ESS}, and \mbox{SSS} on various scalings of PASCAL data.
Figure~\ref{fig:Experiment Results Functions} displays performances of \mbox{SSS} on the \mbox{PASCAL} dataset using stride lengths as functions $f(n)$. Figure~\ref{fig:non-PASCAL} illustrates \mbox{SSS'} performance on RGB color channels of digital art and high-resolution images, alongside correspondingly-sized random matrices. On this diverse set of data, $C$ is used to gain insight into its effects on \mbox{SSS'} accuracy performance. Figure~\ref{fig:IoUs} details \mbox{SSS'} performance across varying IoUs.

\subsection{Discussion}
Figure~\ref{fig:Experiment Results Functions} shows a relative comparison between \mbox{SSS} and \mbox{A-ESS} with \mbox{SSS} using varying stride lengths $f(n)$ on various scalings of PASCAL data. As expected, accuracy and running time decrease as stride lengths increase, offering flexibility by choice of $f(n)$. The relative performance of \mbox{SSS} improves as the input size grows, demonstrating that \mbox{SSS} generally scales more gracefully than \mbox{A-ESS}. In particular, \mbox{SSS} achieves 95\% the accuracy of \mbox{A-ESS} 2x faster at 50\% IoU, the same accuracy 4x faster at 4x scaling, and, at 9x scaling, accuracies up to 99.9\% at 90\% IoU 2.2x faster. We found that \mbox{SSS} performs equally well across all 10 object classes in the PASCAL dataset.

In Figure~\ref{fig:non-PASCAL}, we observe that \mbox{SSS} easily achieves near-perfect accuracy 3.3x faster on all but small, random matrices. On high-resolution image data, it achieves a 45x speedup at 93\% accuracy, an 11x speedup at 99\% accuracy, and 99.5\% accuracy at a 3.5x speedup. Figure~\ref{fig:IoUs} demonstrates that \mbox{SSS} performs increasingly well on more strict IoUs as spatial coherence and matrix size increase. We see that spatial coherence heavily influences \mbox{SSS'} performance: all of the worst performances have significantly lower coherence scores. Thus, spatial coherence affects \mbox{SSS'} accuracy more than $n$ does, making it the single best predictor of accuracy for all datasets used in this work.

We found that poor accuracy performance on small, random matrices is largely due to reaching the iteration cap which prevents infinite loop cases triggered during \mbox{SSS'} search phase. We found that the chance of this case occurring decreases precipitously as matrices increase in size or coherence. We observed this probability to be 20\%, 11\%, and 4\% for \mbox{$n=512, 1024, 2048$} random matrices, respectively, and 0.7\% uniformly across all sets of non-random matrices. If these cases are omitted, \mbox{SSS'} 50\% IoU accuracy on 512x512 random matrices, for example, increases from 78\% to 91\%. \mbox{SSS'} increase in accuracy on increasingly large random matrices in particular can thus be largely attributed to the prevalence of these cases in small, random matrices.

Overall, we observe in \mbox{SSS} an expected trade-off between speed and accuracy. As the size of the inputs increases, \mbox{SSS} scales favorably with an increasing accuracy-time ratio. On full spatially coherent matrices in particular, \mbox{SSS} achieves significant speedups at extremely competitive accuracy.

\section{Conclusion}
We have proposed \mbox{SSS}, a solution to the max-weight rectangle problem that, on many matrices, runs significantly faster than the state-of-the-art with limited to virtually no accuracy loss. The state-of-the-art, \mbox{A-ESS}, runs in linear time and space (\mbox{$O(n^2)$}), whereas \mbox{SSS} runs in sublinear time and space: \mbox{$O(\frac{n^2}{f(n)})$}. We have shown that \mbox{SSS} provides flexibility in a speed-accuracy trade-off setting, and at an increasingly favorable rate as the size and/or spatial coherence of the input increases. It is thus well-suited for real-time applications and against many matrices in general.

{\small
\bibliographystyle{ieee}
\bibliography{egbib}
}

\end{document}